\documentclass[final]{aipproc}
\layoutstyle{8x11single}

\begin{document}

\title 
[Combined Swift BAT-XRT Lightcurves  ]
{Combined Swift BAT-XRT Lightcurves  }

\classification{????, ???? }
\keywords{GRB GRB GRB } 
\author{P. Veres}{address={ Dept. of Physics of Complex Systems, E\" otv\" os University, H-1117 Budapest, P\'azm\'any P. s. 1/A, Hungary, Dept. of Physics, Bolyai Military University, H-1581 Budapest, POB 15, Hungary}}
\iftrue
\author{Z. Bagoly}{address={ Dept. of Physics of Complex Systems, E\" otv\" os University, H-1117 Budapest, P\'azm\'any P. s. 1/A, Hungary}}
\author{J. Kelemen}{address={ Konkoly Observatory, H-1525 Budapest, POB 67, Hungary}}
\author{I. Horv\'ath}{address={ Dept. of Physics, Bolyai Military University, H-1581 Budapest, POB 15, Hungary}}
\author{L. G. Bal\'azs}{address={ Konkoly Observatory, H-1525 Budapest, POB 67, Hungary}}
\author{A. M\'esz\'aros}{address={ Astronomical Institute of the Charles University, V Hole\v{s}ovi\v{c}k\' ach 2, CZ-180 00 Prague 8, Czech Republic}}
\author{G. Tusn\'ady}{address={ R\'enyi Institute of Mathematics, Hungarian Academy of Sciences, H-1364 Budapest, POB 127, Hungary}}
\fi

\copyrightyear  {2008}

\begin{abstract}
In this paper 
we 
make an attempt to combine the two kinds of data from   the Swift-XRT instrument (windowed timing and photon counting modes) and the    from BAT. A thorough desription of the applied procedure will be given. We apply     various binning techniques to the different data: Bayes blocks, exponential     binning and signal-to-noise type of binning. We present a handful of            lightcurves and some possible applications.
\end{abstract}

\date{\today}

\maketitle

\section{Introduction}

A wide spectral coverage leads to a more complete picture of the GRBs. 
In this paper we 
attempt
to combine two of the Swift's instruments IDEZET SWIFTRE! 
namely the XRT 
and the BAT. We investigate the combined properties of the two bands. In order to achieve this, we have extrapolated the gamma-ray flux into the X-ray band.

\section{Data reduction and binning}
We have two kinds of data: $\gamma$-ray and X-ray. We choose three different approaches to bin the lightcurves: Bayesian method for the gamma-ray data, equal binning in logarithmic coordinates in the case of the windowed timing (WT) XRT data and a signal-to-noise type of binning in the case of photon counting (PC) XRT data. The lightcurves and the spectra were generated using standard \texttt{HEASoft} tools and the most recent calibration database. Initial calibration was made using \texttt{xrtpipeline} and
\texttt {batgrbproduct} pipeline scripts.

\subsection{Gamma-rays}
The results of the primary pipeline processing cotain among others a $64$ ms resolution gamma-ray lightcurve. This was used as the input to the Bayesian block analysis. We delimit our combined $15-150$ keV lightcurve using Bayes blocks as presented in \citet{1998ApJ...504..405S}. We set a large prior for the algorithm so it will stop at an early point (which corresponds to a small number of change points) making sure that we have enough resolution for each of the bins. After deducing the time intervals we have used {\texttt batbinevt} to bin the data and get a raw spectrum (pha) file. The further steps recommended in the BAT analysis thread were also carried out. For each interval, we created the appropriate response matrices and fitted both a power law and a power law with a high-energy cutoff. We use the criterion from \citet{2008ApJS..175..179S} to choose between the two models. If the $\chi^2$ improves by more than $6$ by using a cutoff power law, we use the latter instead of the simple power law model. The next step is to extrapolate the model from the gamma-ray band ($15-150$ keV) into the X-ray band ($0.3$ - $10$ keV).

\subsection{X-rays: Windowed Timing}

The WT mode is active when the count rate of the source is high (over $\sim 10$ counts/s). This means we have a good signal-to-noise ratio and we can bin the counts in equal bins in logarithmic space. We have fitted a spectrum for the whole duration of the WT mode and got a conversion factor from rates to flux. There is a more detailed description of the procedure in the next section.
\subsection{ X-rays: Photon Counting}

For the PC mode we bin our data to have a signal-to-noise ratio of at least $3$. We do this by incrementing the endpoint of our interval in time until the count rate reaches the required level. At this point we store this interval and repeat the procedure until the end of the observing period. This type of binning results in roughly equidistant bins in $\log(time)$. We correct for the pile-up in the detector as described in \cite{2006ApJ...638..920V} for the PC mode. For the WT mode pileup correction is made according to \cite{2006A&A...456..917R}.\\

In the following we wish to convert the count rates to flux. To do this, we divide the cumulative lightcurve in $n$ parts, each with equal number of counts. $n$ is chosen by hand depending on the strength of the afterglow from $2$ to $6$. For each time slice we fit a spectrum to get the count equivalent in erg/cm$^2$. The spectra are fitted with \texttt{Xspec} using individual anciliary response functions and the most recent response function available. We used an absorbed power law model of the form: $$ (\mathrm{wabs}_G) \times (\mathrm {zwabs}_S)\times \mathrm{(power law)}.$$ The absorbing column density (NH) of the Milky Way was taken from \citep{1990ARA&A..28..215D} (denoted here by $\mathrm{wabs}_G$), and, where redshift was known, the source absorbtion was also fitted (denoted here by $\mathrm {zwabs}_S$). If the redshift was unknown $\mathrm {zwabs}_S$ was substituted with a simple absorbing $\mathrm {wabs}_S$ component. The spectra were binned using \texttt{grppha} so all channels had minimum $20$ counts.\\

\citet{2007A&A...469..379E} use the same method, but they get the conversion factor (erg/cm$^2$/s equivalent of $1$ count/s ) by integrating over the entire spectrum (corresponding to $n=1$). We report here that our conversion factors are in good agreement with those in \citet{2007A&A...469..379E} and their related web-page.

\section{Individual events}

Here we present several GRBs to illustrate our method of binning and combining data. The behaviour of the afterglow requires the use of a log scale but several GRBs have significant flux before the trigger (precursors at $t < T_{trigger} = 0$). For this reason we choose to include a shift in the time axis where necessary.


\begin{center}
  \includegraphics[width= {0.9\columnwidth} ]{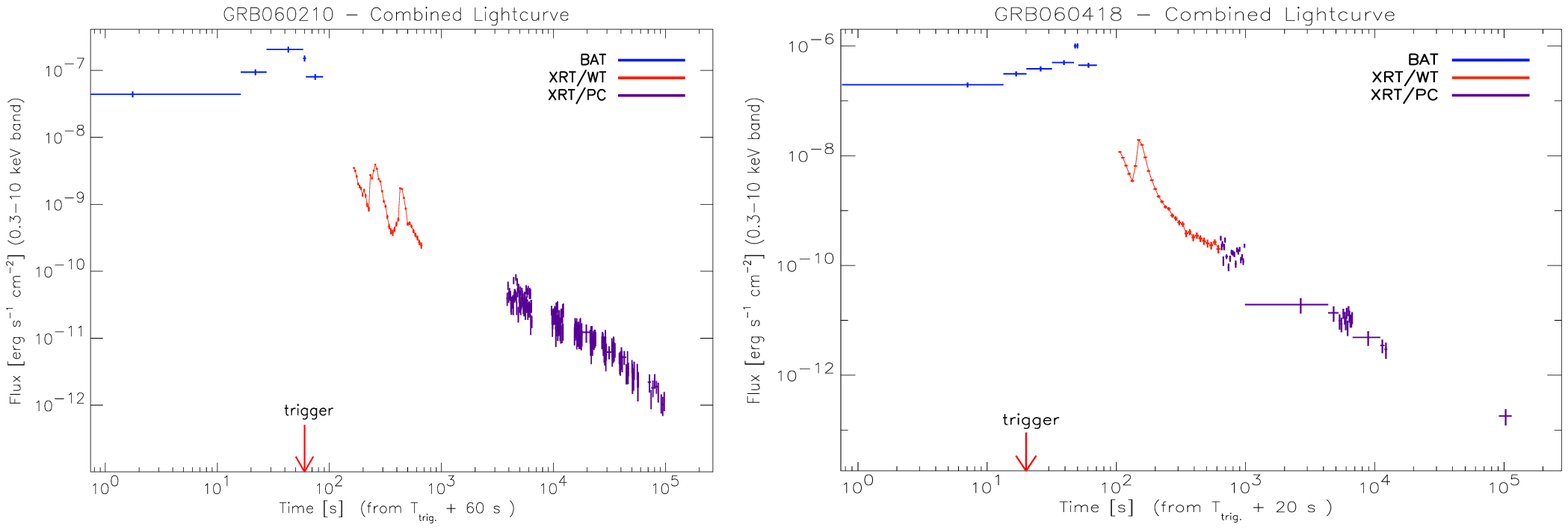}
    \end{center}

\begin{center}
  \includegraphics[width= {0.9\columnwidth} ]{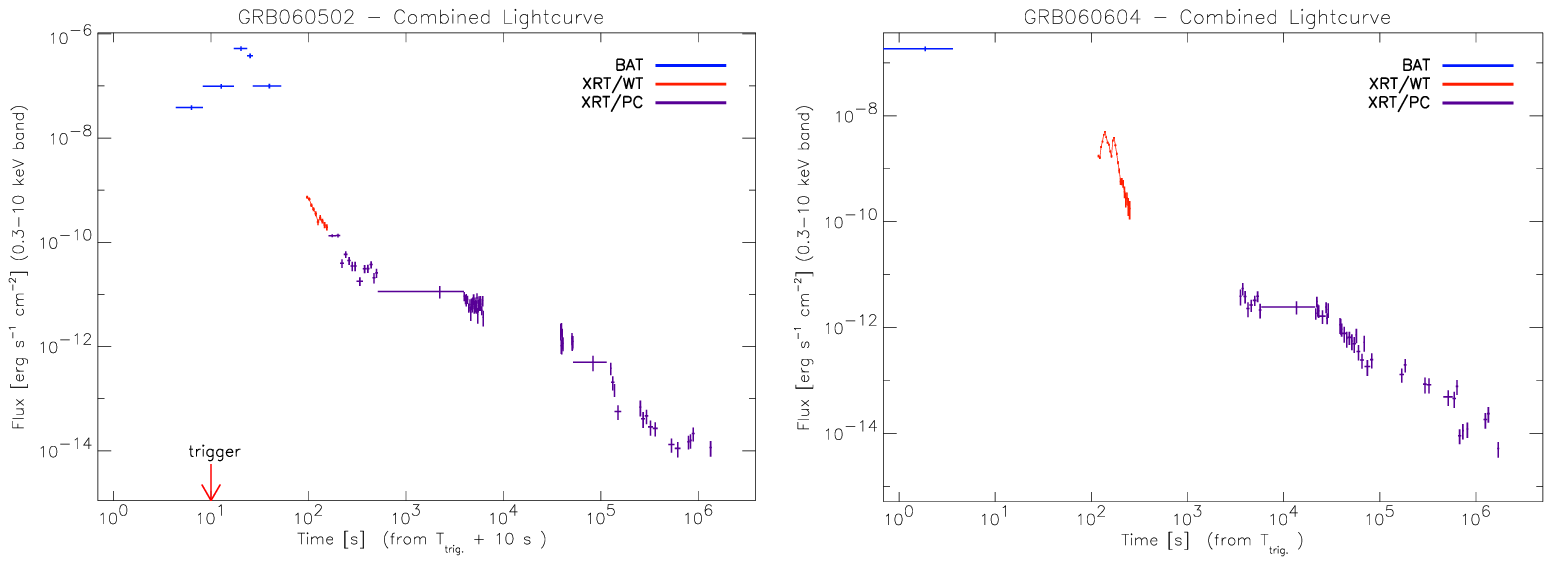}
\end{center}

  
\begin{center}
  \includegraphics[width= {0.9\columnwidth} ]{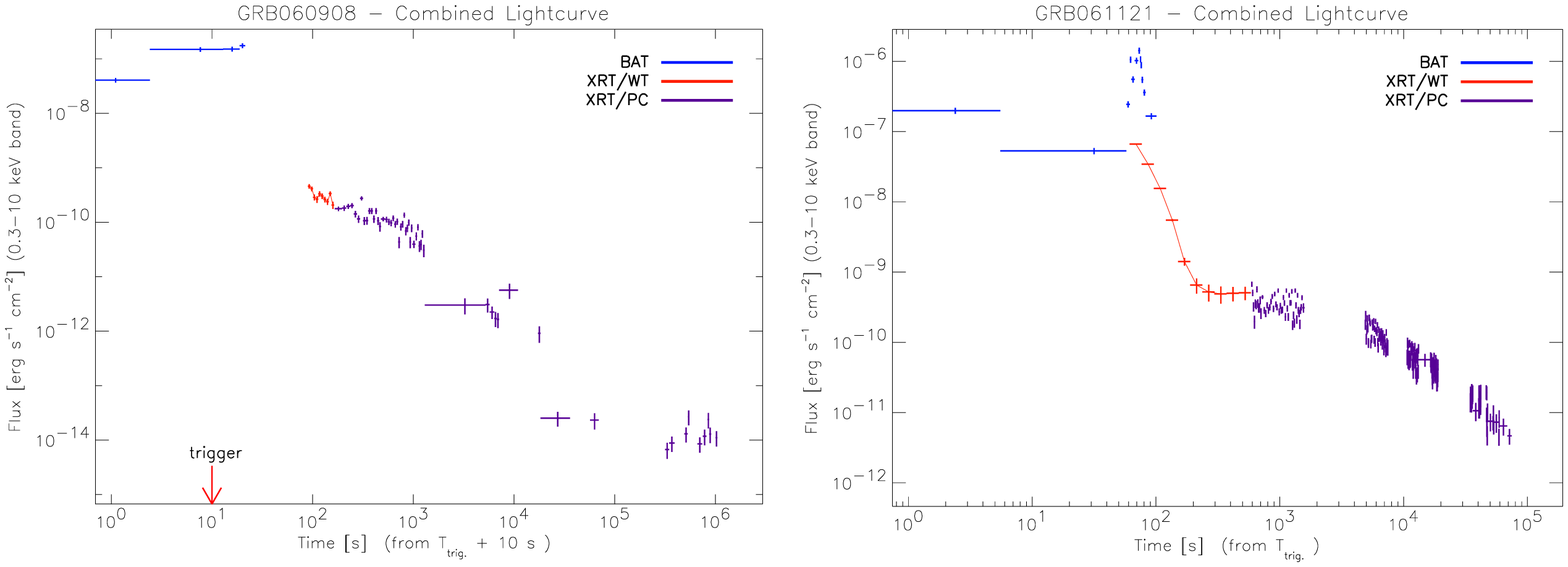}
\end{center}

\begin{center}
  \includegraphics[width= {0.9\columnwidth}]{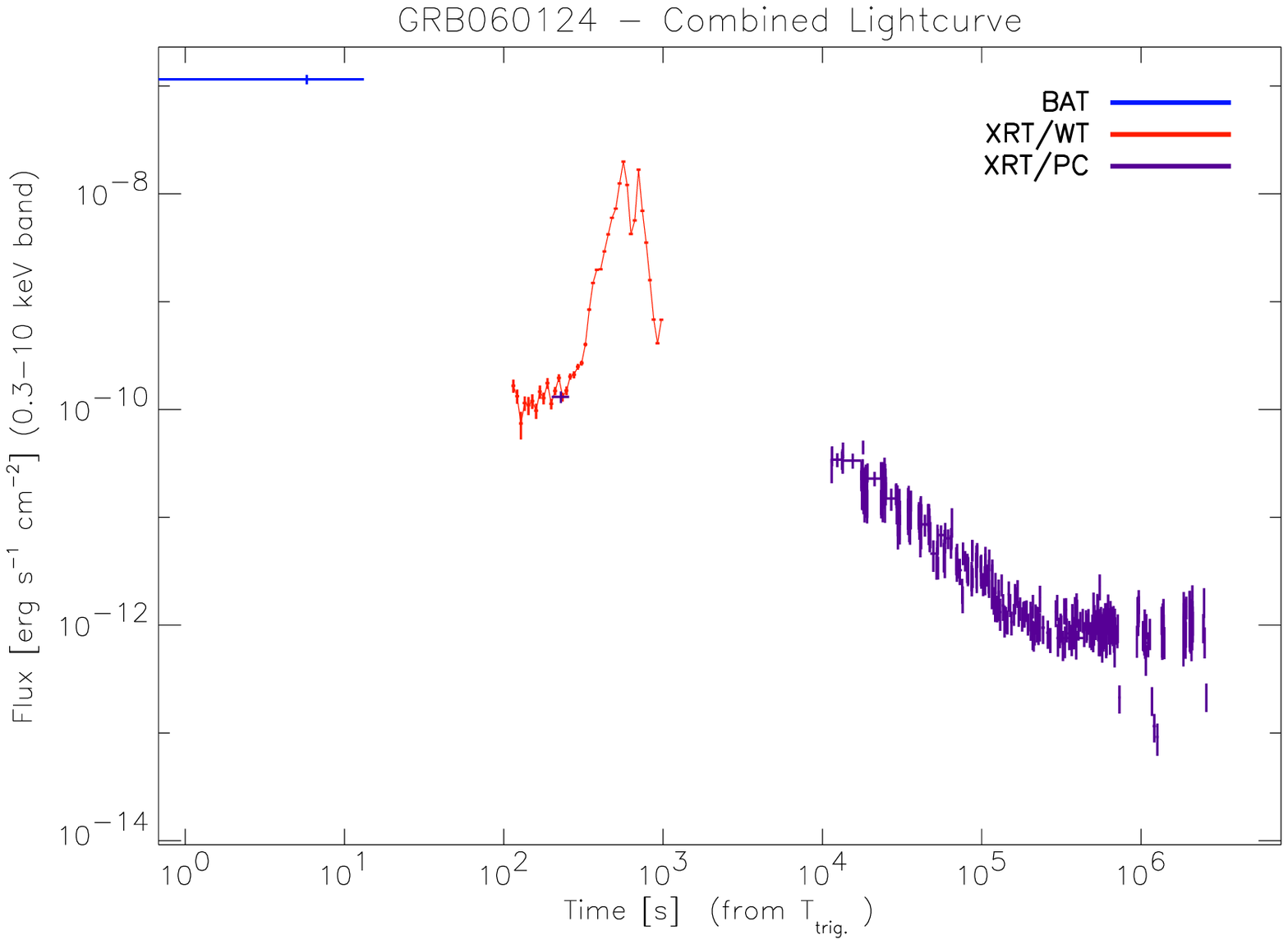}
  
\end{center}

\section{Discussion and Conclusion} 

We have selected a few bright GRBs to present our binning method for data from two bands. There are cases where the possible extrapolation of the WT mode data seems to match the gamma-ray lightcurve (for the first three GRBs: 060210, 060418 and 060502). This could be proof of a connection between the processes which govern the prompt phase and the processes of the early afterglow. It looks possible that the steep decline is indeed a sequel of the prompt phase seen in X-rays. No such claim can be made for the following two events (060604 and 060908) because of the scarceness of the data. At the last two events (061121 and 060124) however, there is an apparent discrepancy between the gamma-ray curve and the X-ray curve. This discrepancy is best seen at GRB061121. This could mean the extrapolation of the gamma-ray spectrum was not adequate.

\begin{theacknowledgments}
This study was supported by the Hungarian OTKA grant No. T48870, by a Research
Program MSM0021620860 of the Ministry of Education of Czech Republic, by a GAUK
grant No. 46307 and by a Bolyai Scholarship (I.H.).
\end{theacknowledgments}


\bibliographystyle{aipproc}
\bibliography{nanking}

\end{document}